\PassOptionsToPackage{x11names}{xcolor}
\documentclass[prd,nofootinbib,preprint,superscriptaddress]{revtex4}

\usepackage{amsmath, amssymb, amsthm, graphicx, epsfig, fancyhdr,epsfig, slashed, mathrsfs}
\usepackage{subfigure}
\usepackage{tikzsymbols}
\usepackage{natbib}
\usepackage{float}

\usepackage{physics}

\usepackage{mathtools} 

\usepackage{tikz,hyperref}

\usepackage{amsfonts}
\usepackage{mathtools}

\newcommand{\mpl}{M_{\text{Pl}}}

\newcommand{\be}{\begin{equation}}
\newcommand{\ee}{\end{equation}}

\newcommand{\bea}{\begin{eqnarray}}
\newcommand{\eea}{\end{eqnarray}}
\newcommand{\beq}{\begin{eqnarray}}
\newcommand{\eeq}{\end{eqnarray}}

\newcommand{\Rmnum}[1]{\expandafter\@slowromancap\romannumeral #1@}

\newcommand{\tv}{\bar{v}}

\definecolor{lime}{HTML}{A6CE39}
\DeclareRobustCommand{\orcidicon}{
	\begin{tikzpicture}
	\draw[lime, fill=lime] (0,0) 
	circle [radius=0.2] 
	node[white] {{\fontfamily{qag}\selectfont \tiny ID}};
	\draw[white, fill=white] (-0.0625,0.095) 
	circle [radius=0.007];
	\end{tikzpicture}
	\hspace{-2mm}
}

\foreach \x in {A, ..., Z}{\expandafter\xdef\csname orcid\x\endcsname{\noexpand\href{https://orcid.org/\csname orcidauthor\x\endcsname}
			{\noexpand\orcidicon}}
}

\allowdisplaybreaks

\begin{document}

\title{Imprints of a Supercooled Phase Transition in the Gravitational\\ Wave Spectrum from a Cosmic String network}

\author{Francesc Ferrer\orcidF{}}
\email{ferrer@wustl.edu}
\affiliation{Department of Physics and McDonnel Center for the Space Sciences, \\
Washington University, St. Louis, MO 63130, USA}

\author{Anish Ghoshal\orcidA{}}
\email{anish.ghoshal@fuw.edu.pl}

\author{Marek Lewicki\orcidC{}}
\email{marek.lewicki@fuw.edu.pl}
\affiliation{Institute of Theoretical Physics, Faculty of Physics, University of Warsaw, \\ ul. Pasteura 5, 02-093 Warsaw, Poland}

\begin{abstract}
A network of cosmic strings (CS), if present, would continue emitting gravitational waves (GW) as it evolves throughout the history of the Universe. This results in a characteristic broad spectrum making it a perfect source to infer the expansion history. 
In particular, a short inflationary period caused by a supercooled phase transition would cause a drop in the spectrum at frequencies corresponding to that event.
However, the impact on the spectrum is similar to the ones caused by an early matter-dominated era or from particle production, making it difficult to disentangle these different physical origins. We point out that, in the case of a short inflationary period, 
the GW spectrum receives an additional contribution from the phase transition itself. This leads to a characteristic imprint of a peak on top of a wide plateau both visible at future GW observatories.
\end{abstract}

\maketitle

\section{Introduction}
\label{sec:intro}

Both the Standard Model of Particle Physics (SM) and the Standard $\Lambda$CDM
Cosmological Model describe with remarkable accuracy a vast amount of 
observational data on the fundamental interactions of
elementary particles~\cite{ParticleDataGroup:2022pth} and on the dynamics
and composition of the Universe~\cite{Planck:2018vyg}. Nevertheless, 
given the large number of unexplained free parameters, this framework is 
better viewed as an effective description, one that is incomplete at the 
fundamental level. 
For instance, explaining the matter-antimatter asymmetry or the nature of
the dark matter requires extending the SM, possibly with new particles 
featuring additional interactions. These additional degrees of freedom
might induce deviations from the standard early universe 
evolution~\cite{Allahverdi:2020bys}. Well known examples are an 
early matter dominated period before the universe became radiation dominated, 
or a kination era when the background equation-of-state is more stiff 
than that of radiation~\cite{Peebles:1998qn,Poulin:2018dzj}.

In this respect, a completely new avenue towards the observation of our 
universe was recently opened with the first detection of GWs by the LIGO/Virgo 
collaboration~\cite{LIGOScientific:2016aoc,LIGOScientific:2018mvr,LIGOScientific:2020ibl,LIGOScientific:2021usb,LIGOScientific:2021djp}. 
This experimental breakthrough, in turn, 
has helped to accelerate the development of many planned future GW 
observatories with vastly improved sensitivities~\cite{Punturo:2010zz,Hild:2010id,Janssen:2014dka,Graham:2016plp,LISA:2017pwj,Graham:2017pmn,Badurina:2019hst,AEDGE:2019nxb,Bertoldi:2021rqk,Alonso:2022oot,Badurina:2021rgt}.
Furthermore, while all confirmed detections to date are associated
with transient 
astrophysical sources, 
hope rises to probe the stochastic backgrounds of gravitational waves (SGWB) 
that may be generated by 
early Universe processes~\cite{LISACosmologyWorkingGroup:2022jok}.
Importantly, the observation of a SGWB originating in the early Universe 
would enable us to explore the history of the universe before the epoch of 
Big-Bang Nucleosynthesis (BBN) and, 
therefore, to study new physics at high energies beyond the reach 
of particle physics laboratory-based 
experiments~\cite{LISACosmologyWorkingGroup:2022jok}.
Nevertheless, we should keep in mind that 
the populations of compact objects originating the currently observed events
will also produce their own GW foregrounds making the detection of 
any primordial component more challenging~\cite{Lewicki:2021kmu}. 

Another result that has energised the theoretical community is the recent observation by pulsar-timing arrays 
(PTAs)~\cite{NANOGrav:2020bcs,Goncharov:2021oub,Chen:2021rqp,Antoniadis:2022pcn} of a common spectrum process which could be the first indicator of the 
upcoming detection of a stochastic GW background. 
This background could potentially be of primordial origin~\cite{Vaskonen:2020lbd,DeLuca:2020agl,Nakai:2020oit,Ratzinger:2020koh,Kohri:2020qqd,Vagnozzi:2020gtf,Neronov:2020qrl,Blanco-Pillado:2021ygr,Wang:2022wwj,RoperPol:2022iel,Ferreira:2022zzo} with the foremost candidate being a network of cosmic strings that is
the focus of the present work~\cite{Ellis:2020ena,Blasi:2020mfx}. 
the latest PTA data that has been recently released shows strong evidence that the signal is caused by GWs~\cite{NANOGrav:2023gor,NANOGrav:2023hde,Antoniadis:2023ott,Antoniadis:2023lym,Zic:2023gta,Reardon:2023gzh,Xu:2023wog}. Nevertheless, standard cosmic strings provide a poor fit to the observed frequency dependence of the signal, which is better described by e.g. models with smaller string intercommutation probability~\cite{Ellis:2023tsl,NANOGrav:2023hfp, Antoniadis:2023zhi}. Fortunately, the interplay between supercooling and string network dynamics that we will discuss in this paper carries over directly to these generalized scenarios~\cite{Ellis:2023tsl}.

The fact that the SGWB produced by cosmic strings in standard cosmology is 
scale-invariant~\cite{Vilenkin:1981bx, Hogan:1984is, Vachaspati:1984gt, Accetta:1988bg, Bennett:1990ry, Caldwell:1991jj, Allen:1991bk, Battye:1997ji, DePies:2007bm, Siemens:2006yp, Olmez:2010bi, Regimbau:2011bm, Sanidas:2012ee, Sanidas:2012tf, Binetruy:2012ze, Kuroyanagi:2012wm, Kuroyanagi:2012jf,Sousa:2016ggw,Sousa:2020sxs} makes it an excellent tool to probe the 
expansion of the early Universe~\cite{Cui:2017ufi,Cui:2018rwi,Auclair:2019wcv, Guedes:2018afo, Ramberg:2019dgi, Gouttenoire:2019kij, Gouttenoire:2019rtn,Chang:2019mza,Cui:2019kkd, Gouttenoire:2021jhk}, since deviations from the 
standard evolution would leave characteristic imprints on the otherwise
largely featureless SGWB.

An important example is provided by very strong phase transitions leading to
a period of supercooling, as expected in particle physics models 
involving confinement~\cite{Creminelli:2001th, Randall:2006py,Nardini:2007me,Konstandin:2011dr,vonHarling:2017yew,Iso:2017uuu,Bruggisser:2018mrt,Baratella:2018pxi,Agashe:2019lhy,DelleRose:2019pgi,vonHarling:2019gme,Caprini:2019egz,Baldes:2020kam,Sagunski:2023ynd} or quasi conformal 
models ~\cite{Jinno:2016knw,Marzola:2017jzl,Prokopec:2018tnq,Marzo:2018nov,VonHarling:2019rgb,Aoki:2019mlt,Wang:2020jrd,Ellis:2020nnr,Ghoshal:2020vud,Lewicki:2021xku,Dasgupta:2022isg,Kierkla:2022odc,Wong:2023qon, Salvio:2023qgb}.
Supercooling causes a significant modification of the expansion rate as the field undergoing the transition is trapped in the false vacuum state whose energy dominates the expansion, thus leading to a new short period of thermal 
inflation. 
If a network of cosmic strings were present during the phase transition,
this period of accelerated expansion would cause a dip in the strings'
SGWB spectrum. Such a feature, however, is similar to the changes that 
would be caused by an epoch of early matter
domination~\cite{Cui:2018rwi,Guedes:2018afo,Gouttenoire:2019kij}, which makes
it difficult to disentangle these different physical origins.
On the other hand, it should be emphasized that supercooled phase transitions 
are themselves very strong sources of GWs. 
In this paper, we propose a smoking gun signature of a period of supercooling utilising the SGWB from local cosmic strings. 
Its uniqueness comes from our ability to observe both the GW background 
produced by the phase transition at the end of the supercooling period as well as the effect that it would have on the spectrum generated by the network of cosmic strings.

\textit{The paper is organized as follows:} In section~\ref{sec:Supercooled_PT_and_thermal_inflation} we define the basic connection between the phase transition (PT) parameters and the length of the thermal inflation period. In sec.~\ref{sec:GW_from_strings} we outline the computation of the GW spectrum produced by cosmic strings and the impact that a period of supercooling will have on it. Section~\ref{sec:GW_from_PT} is devoted to the signal produced by the PT at the end of the supercooling period. Finally, in sec.~\ref{sec:results} we discuss the results and we conclude in sec.~\ref{sec:conclusion}.


\section{Supercooled PT and thermal inflation}
\label{sec:Supercooled_PT_and_thermal_inflation}

Our first goal is  to establish the connection between the parameters 
describing a first order PT and the length of the inflationary epoch it 
causes. 
The first key parameter of the transition is its strength which in the case of strong transitions can be defined as the ratio of vacuum energy density to the radiation background~\cite{Ellis:2018mja}
\be \label{eq:alpha}
\alpha = \rho_{V}/\rho_{R*},
\ee
where the subscript $*$ indicates the time of the transition.
We will be mostly interested in cases where the vacuum energy dominates the expansion for a significant amount of time. The length of the inflationary period
is measured by the usual number of efolds,
\be \label{eq:Ne}
N_e=\log \left(a_*/a_V\right),
\ee
where the subscript $V$ refers to the moment when the vacuum energy begins to dominate the expansion. Using the fact that $\rho_R\propto a^{-4}$ while $\rho_V$ remains constant we can easily relate eq.~\ref{eq:alpha} and eq.~\ref{eq:Ne}:
\be \label{eq:alpha_Ne}
\alpha=\text{e}^{4N_e} \, ,
\ee
which provides the link between the strength of the transition and the number of e-folds of inflation it causes.

The second key parameter describing the transition is the temperature reached as the universe is reheated $T_\text{reh}$. 
We will assume instantaneous reheating for simplicity, meaning the 
temperature can be related to the vacuum energy density in the usual way $T_\text{reh} \propto \rho_V^{1/4}$. However, in practice, we will keep $T_\text{reh}$ as a free parameter and, given that it is the same as the temperature when the period of supercooling began, it can also be understood as the moment when the evolution first became non-standard. 
On the other hand, the strength $\alpha$ of the preceding transition (see eq.~\eqref{eq:alpha_Ne}) dictates the duration of that non-standard period.

Finally, one last important parameter characterizing the transition is its time scale $\beta$, which is usually introduced through an approximation of the vacuum decay rate
\be
\Gamma \propto \text{e}^{\beta(t-t_*)}\, .
\ee
The value of $\beta$ is usually computed by differentiating the decay rate 
calculated in the particular model under consideration. More generally, in 
classically scale invariant models featuring such strong 
transitions~\cite{Ellis:2019oqb,Ellis:2020nnr} this parameter roughly 
asymptotes to $\beta/H \approx 10$ as the strength of the transition grows~\cite{Ellis:2020awk}. We take this as our benchmark value.

\section{Effects of supercooling on the gravitational wave emission from a cosmic string network}
\label{sec:GW_from_strings}

Scenarios of physics beyond the standard model commonly predict the existence of additional global or local symmetries. The spontaneous breaking of such  symmetries in the early universe might lead to the formation of a network of cosmic strings via the Kibble 
mechanism~\cite{Kibble:1976sj,Kibble:1980mv,Hindmarsh:1994re,Vilenkin:2000jqa}.

Numerical simulations have established that, as long as the scale factor of 
the universe grows as a power law, a network of cosmic strings reaches a 
scaling regime~\cite{Vanchurin:2005pa,Martins:2005es,Blanco-Pillado:2011egf},
characterized by a constant mean velocity of long strings $\bar{v}$
and a characteristic correlation length $L$ that remains constant with
respect to the Hubble horizon $d_H$. 
The correlation length is defined by~\cite{Vilenkin:2000jqa}:
\beq
\rho_{\infty} ~\equiv~ \frac{\mu}{L^2} \ ,
\eeq
where $\mu$ is the string tension, and $\rho_\infty$ is the energy density
of long strings in the network, which form a Brownian random walk on large
scales~\cite{Scherrer:1986sw}. As the universe expands, the correlation length stretches with the scale factor, which tends to increase the energy density of long strings compared to the radiation background. But this increase
is compensated by energy losses associated with the production of small loops
that form when long strings cross each other, and the network evolves towards
the scaling regime~\cite{Vachaspati:1984gt}. 
Once formed, the loops decouple
from the network and start to decay. For strings generated from the 
breaking of a local symmetry, which we focus on, the decay proceeds primarily 
through the emission of gravitational waves. 
When the expansion of the universe changes, e.g. from a radiation dominated 
to a matter dominated epoch, the network enters a transient regime until
it settles into the new scaling regime. 

The
evolution of the long-string network borne out by simulations is well 
approximated by the velocity-dependent one-scale~(VOS) model~\cite{Martins:1995tg,Martins:1996jp,Martins:2000cs,Avelino:2012qy,Sousa:2013aaa}. In this
model, the parameters $L$ and $\bar{v}$ that characterize the string network
evolve according to~\cite{Martins:1996jp,Martins:2000cs}:
\beq
\dv{L}{t} &=& (1+\tv^2)\,HL + \frac{\tilde{c}\tv}{2}
\label{eq:vos1}\\
\dv{\tv}{t} &=& (1-\tv^2)
\left[\frac{k(\tv)}{L} - 2H\,\tv\right] \ ,
\label{eq:vos2}
\eeq
where the momentum parameter
\beq \label{eq:vos3}
k(\tv) = \frac{2\sqrt{2}}{\pi}(1-\tv^2)(1+2\sqrt{2}\tv^3)
\left(\frac{1-8\tv^6}{1+8\tv^6}\right) \ ,
\eeq
measures the deviation from straight strings with $k(\bar{v})=1$,
and $\tilde{c} \simeq 0.23$ describes closed loop 
formation~\cite{Martins:2000cs}.
It is well known that scale invariant solutions of the form $L = \xi t$
(i.e. $L \propto 1/H$) and constant $\bar{v}$ exist for the 
system~(\ref{eq:vos2}) when the scale factor is a power law. 

On the other hand, during the epoch of supercooling, the strings are 
exponentially stretched and the network is diluted due to the accelerated
expansion of the universe~\cite{Guedes:2018afo}. 
The network departs from the scaling regime
and the production of loops is suppressed. We 
follow ref.~\cite{Cui:2019kkd}, and we use a simplified picture of 
supercooling and secondary reheating together with the VOS model 
to estimate the dilution of cosmic strings and their subsequent evolution.

More specifically, we assume at early times radiation domination $\rho\propto a^{-4}$ up until the Hubble parameter reaches $H_{V} = \Delta V/3\mpl^2 \,$. The energy density of the universe is then dominated by vacuum energy while the scale factor grows by a fixed number of efolds $N_e$ (see eq.~\eqref{eq:Ne}) until the phase transition is completed and standard radiation domination resumes.
The energy scale during this period is fixed by our choice of the reheating 
temperature $T_\text{reh} = (30/\pi^2g_* \Delta V)^{1/4} $, assuming for 
simplicity that the reheating is instantaneous.
In this inflating background, we track the evolution of the network using 
the VOS equations~(\ref{eq:vos1}-\ref{eq:vos3}) beginning with the standard 
scaling initial conditions in the early radiation dominated period before 
supercooling sets in. During the supercooling period, the network is diluted, with the correlation length growing beyond the Hubble scale $HL \gg 1$ and 
the velocity dropping $\bar{v}\ll 1$. 
Once radiation domination resumes, the correlation length follows 
$L\propto a$ as the network energy regrows until it eventually 
reaches again the scaling solution. In contrast to the case of strings 
diluted by inflation, here we expect that the string network will always grow 
back to scaling relatively quickly and the low frequency part of the spectrum 
will remain unchanged~\cite{Cui:2019kkd}.

The network of cosmic strings acts as a long-lasting source of gravitational 
waves from the time of their production until today. The emission is 
dominated by the closed string loops that are formed when long strings in the 
network intersect~\cite{Vachaspati:1984gt}. Gravitational waves emitted by
oscillating loops at different epochs generate a SGWB over a large frequency
range, and events such as phase transitions leave imprints on the resulting GW 
spectrum. This makes the SGWB an invaluable tool to gain insight into 
modifications of the expansion of the universe induced by high-energy physics
effects~\cite{Cui:2017ufi,Cui:2018rwi}.

To determine the SGWB from the cumulative emission of closed loops we note
that recent simulations of Nambu-Goto string networks observe that long strings can be separated into a population of large and  non-relativistic loops, with initial size $l_i = \alpha_L L \pqty{t_i}$ 
and $\alpha_L \approx 0.37$ such that in radiation domination we have $l_i \approx \alpha t_i$ with $\alpha\approx 0.1$, as well as a collection of smaller and highly relativistic loops. The energy in the smaller loops is mostly in the form of kinetic energy that redshifts away. Hence, large loops are the main source of the SGWB. Roughly a fraction ${\cal F}_\alpha \sim 0.1$ of the total energy is transferred into loops~\cite{Blanco-Pillado:2013qja,Blanco-Pillado:2015ana,Blanco-Pillado:2017oxo,Blanco-Pillado:2019vcs,Blanco-Pillado:2019tbi},
and we include an additional reduction factor $f_r=\sqrt{2}$ that accounts for
the fraction of energy lost into peculiar velocities of loops~\cite{Auclair:2019wcv}.
 After their formation at time $t_i$, a loop will oscillate and gradually 
 shorten due to energy losses to GWs:
\begin{equation}\label{eq:loopsize}
	l (t) = \alpha_L L \pqty{t_i} - \Gamma G \mu \pqty{t -t_i},
\end{equation}
where $\Gamma \approx 50$ is the total rate of GW 
emission~\cite{Blanco-Pillado:2017oxo,Blanco-Pillado:2017rnf}. 
To compute the spectrum we will need to sum the emission from all normal modes of a closed loop.
We will start from the fundamental one whose total emission can be expressed as~\cite{Avelino:2012qy,Auclair:2019wcv}:
\beq
\label{eq:GWdensity2}
\Omega_{\rm GW}^{(1)}(f) =
\frac{16\pi}{3 f}
\frac{\mathcal{F}_{\alpha}}{f_r}
\frac{G\mu^2 }{H_0^2}
\frac{ \Gamma}{\alpha_L \zeta(q)}
\int_{t_F}^{t_0}\!d\tilde{t}\;
\frac{1}{\alpha_L \dot{L}(t_i)+\Gamma G\mu}
\frac{\tilde{c} v(t_i)}{L(t_i)^4} 
\bigg[\frac{a(\tilde{t})}{a(t_0)}\bigg]^5
\bigg[\frac{a(t_i)}{a(\tilde{t})}\bigg]^3
\,\Theta(t_i - t_F)
\eeq
where we integrate over the emission time $\tilde{t}$, and the time of formation of contributing loops $t_i$ is found using Eq.~\eqref{eq:loopsize} taking into account that loops with size $l(\tilde{t},f)=\frac{2}{f}\frac{a(\tilde{t})}{a(t_0)}$ emit at frequency $f$. We also added $\zeta(q)=\sum_k q^{-q}$ which assuming emission by cusps gives $\zeta(4/3)\approx 3.6$  and ensures the power in all modes sums to $\Gamma$. 
Finally, $t_F$ is the initial time at which the network first reached scaling after its formation.

Assuming the emission is dominated by cusps, the contribution from 
higher modes can be expressed through that the first one:
\begin{equation}
\Omega_{\rm GW}^{(k)}(f) =k^{-4/3}\,\Omega_{\rm GW}^{(1)}(f/k) \, .
\end{equation}
The total abundance is then a sum of the emission from all modes,
which we approximate as: 
 \beq \label{eq:modesum}
 \Omega_{GW}(f) \simeq \sum_{k=1}^N\Omega_{GW}^{(k)}
 + \int_{{N+1}}^{\infty}\!dk\;\Omega_{GW}^{(k)} \ ,
 \eeq
 where above $N=10^3$ we approximate the sum with an integral which leads to accurate predictions with no significant impact on the computation time~\cite{Cui:2019kkd}.

\section{Gravitational Waves from Phase Transitions}
\label{sec:GW_from_PT}

There are several sources of GWs associated with a phase transition that have been discussed in the literature~\cite{Caprini:2019egz,Caprini:2015zlo}. 
Firstly the bubble wall collisions~\cite{Kosowsky:1992vn,Huber:2008hg,Lewicki:2019gmv,Lewicki:2020azd} featured in extremely supercooled scenarios~~\cite{Ellis:2018mja,Ellis:2019oqb,Lewicki:2019gmv,Lewicki:2020jiv,Lewicki:2020azd}. 
Secondly, the motion of the fluid shells propelled by the growing bubbles~\cite{Kamionkowski:1993fg,Hindmarsh:2013xza,Hindmarsh:2015qta,Hindmarsh:2017gnf,Hindmarsh:2020hop}. Finally a possible contribution from turbulence produced as the plasma flow becomes non-linear~\cite{RoperPol:2019wvy,Kahniashvili:2020jgm,Pol:2021uol,Auclair:2022jod}.

Since we focus exclusively on strongly supercooled phase transitions, we can safely assume that the wall velocity is close to the speed of light,
as this would be 
true even for much weaker transitions~\cite{Laurent:2020gpg,Cline:2021iff,Lewicki:2021pgr,Laurent:2022jrs,Ellis:2022lft}. Assuming that the strength of the 
transition is large $\alpha \gg 1$, we do not need to diligently compute the energy budget~\cite{Espinosa:2010hh,Ellis:2019oqb,Ellis:2020nnr} in order to check what fraction of energy will be deposited into each of the sources. 
This is because in this regime the fluid shells propelled by the bubbles behave as relativistic shocks~\cite{Jinno:2019jhi}. 
These are very peaked and continue propagating at relativistic velocities.
Their energy also dissipates at the same rate as that of the scalar field gradients after the collision and, as a result, the GW spectrum they produce is
identical to the one produced by bubble wall collisions~\cite{Lewicki:2022pdb}.      
The spectrum takes the form~\cite{Lewicki:2020azd,Lewicki:2022pdb}: 
\bea
&\Omega_{\rm GW}^{\rm PT}(f)h^2 = 1.67\times 10^{-5}\left( \frac{100}{g_*(T_{\text{reh}})} \right)^\frac13 \left(\frac{\beta}{H}\right)^{-2} \!\left(\frac{ \alpha}{1+\alpha}\right)^2 \!\! \frac{A\,(a+b)^c}{\left[b \!\left(\frac{f}{f_p}\right)^{\!\text{-}\frac{a}{c}} \!+ a \!\left(\frac{f}{f_p}\right)^{\!\frac{b}{c}}\right]^c} \,,
\label{eq:PTabundance}
\eea
where, $a=b=2.4$, $c=4.0$ and $A=5.13\times 10^{-2}$\, .
The peak frequency is given by
\begin{equation}
	f_{\rm p} \,=1.65\times 10^{-5} {\rm Hz} \left(\frac{T_{\text{reh}}}{100 {\rm GeV}}\right)\left(\frac{g_*}{100}\right)^\frac16 \frac{0.71}{2\pi} \left( \frac{\beta}{H} \right) \, ,
\end{equation}
and $g_*$ is the number of relativistic degrees of freedom that we 
approximate using the results of~\cite{Saikawa:2018rcs}.

The signal-to-noise ratio we refer to in the results that follow 
is computed in the standard way
\begin{equation} \label{eq:SNR}
	\text{SNR} \equiv  \left[ \tau  \int_{f_{min}}^{f_{max}} \left( \frac{\Omega_{\text{sw}}(f)}{\Omega_{\text{noise}}(f)}  \right)^2  \dd{f}
	\right]^{1/2} \, ,
\end{equation}
where we take the noise curve for a given experiment and assume the duration of each mission to be $\tau=4$~years.

\section{Resulting GW spectra}

Finally, we can put together the GW spectra computed for cosmic strings in sec.~\ref{sec:GW_from_strings} and from the phase transition ending the supercooling inflationary period discussed in sec.~\ref{sec:GW_from_PT}.
We contrast these with the predicted sensitivities of LISA~\cite{Bartolo:2016ami,Caprini:2019pxz}, SKA~\cite{Janssen:2014dka}, AION-1km~\cite{Badurina:2019hst},
AEDGE~\cite{Bertoldi:2019tck}, AEDGE+~\cite{Badurina:2021rgt}, 
and ET~\cite{Punturo:2010zz,Hild:2010id} experiments, as well as the currently running 
LIGO/Virgo/KAGRA~\cite{TheLIGOScientific:2014jea,Thrane:2013oya,TheLIGOScientific:2016wyq,LIGOScientific:2019vic}.

\begin{figure}[t]
    \centering
  
\includegraphics[width=0.49\linewidth]{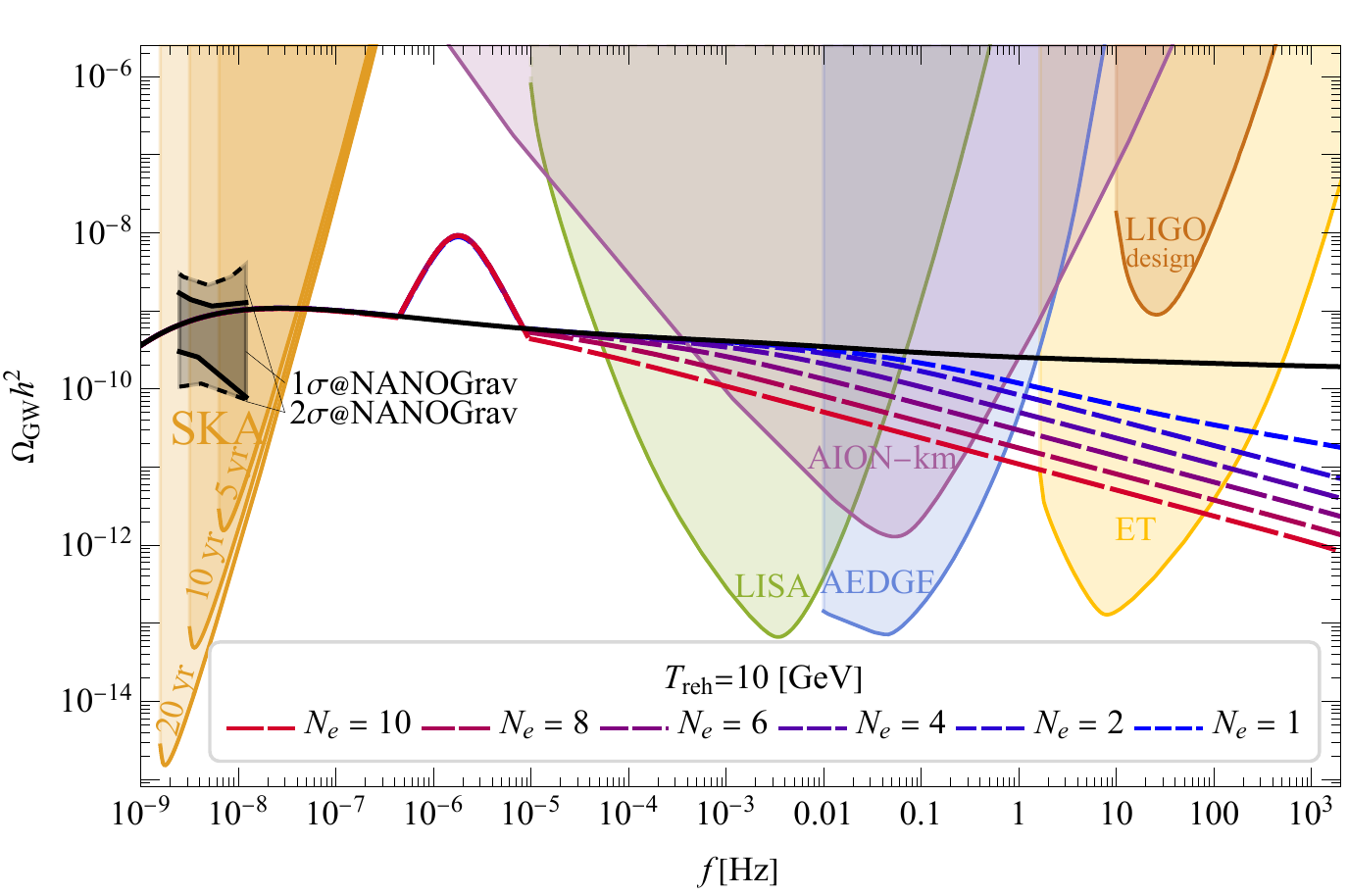}
\includegraphics[width=0.49\linewidth]{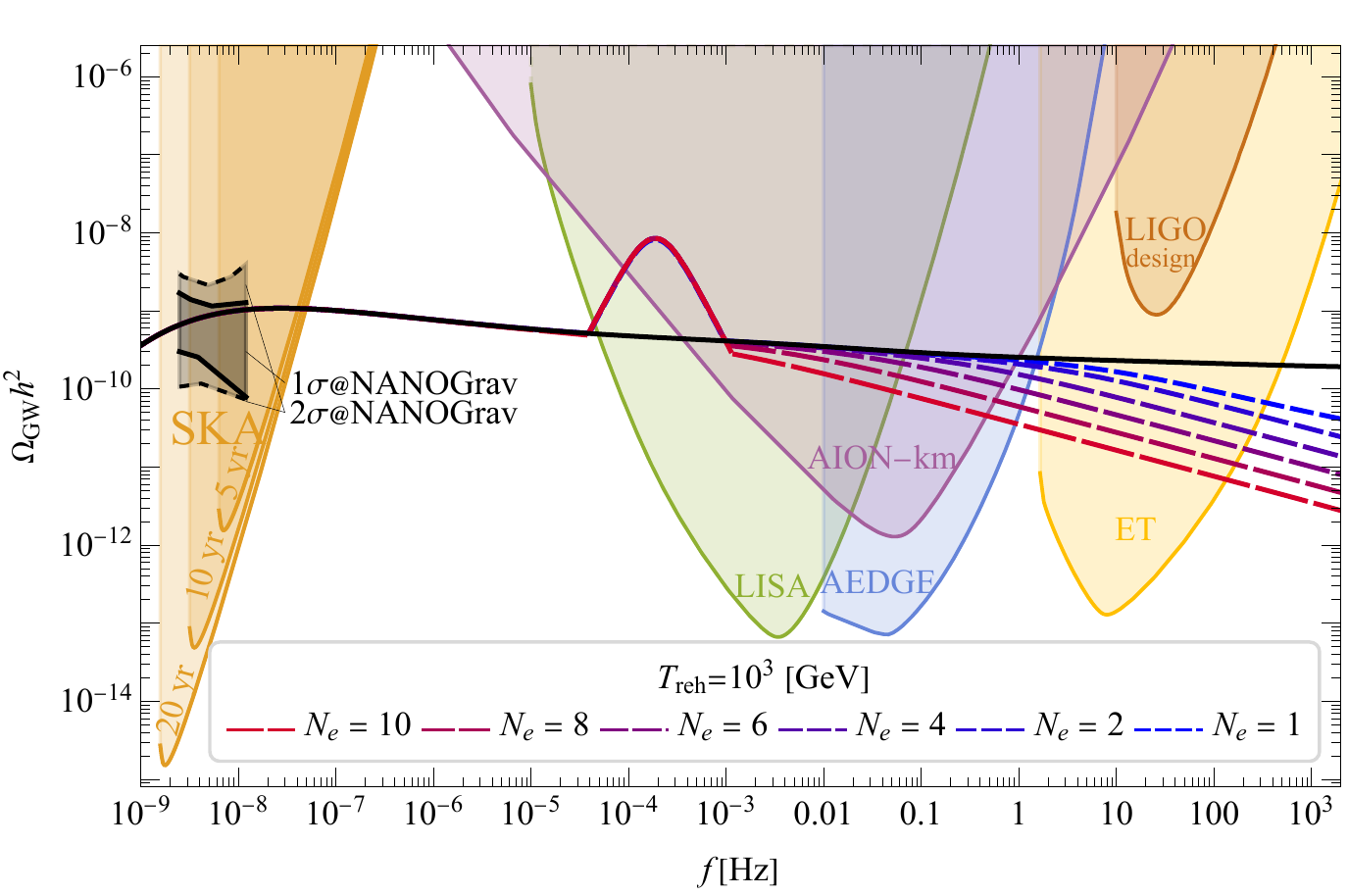}
    \caption{\it \label{fig:SpectraPlot} \footnotesize GW spectra from a supercooled phase transition and cosmic strings. The thick black line shows the cosmic string signal with no supercooled phase transition, while the coloured dashed lines show both sources for the number of efolds of supercooling indicated. The bump above the black line is the contribution from the transition itself, while the falling lines at higher frequencies illustrate how the string spectrum is impacted by the short inflation period associated with supercooling. For the strings, we used $G\mu=10^{-10}$ while for the PT $\beta/H=10$ in this example.}
\end{figure}

Figure~\ref{fig:SpectraPlot} shows the combined spectra. In all the panels we set the tension of the strings to $G\mu=10^{-10}$ inspired by the NANOGrav data~\cite{Ellis:2020ena} and the time scale of the phase transition is $\beta/H=10$ as a benchmark for supercooled scenarios~\cite{Ellis:2020awk}. The solid black line is the cosmic string spectrum with no period of supercooling, while the increasingly red lines with longer dashing show the combined spectra with a period of supercooling from $N_e=0.5$ to $N_e=12$ efolds, which covers what is 
achievable in typical models~\cite{Lewicki:2021xku,Konstandin:2011dr}. 
The left panel shows the results for reheating temperature of $T_{\text{reh}}=10$ GeV while the right one $T_{\text{reh}}=10^{3}$ GeV. The temperature controls the frequency of the features in the spectrum as both the PT signal~\cite{Caprini:2015zlo,Caprini:2019egz} and the frequency at which the cosmic string plateau is modified~\cite{Cui:2017ufi,Cui:2018rwi} grow linearly with temperature. 
The modified string spectra at high frequency follow a $f^{-\frac13}$ power-law.
The peak above the string spectrum comes from the phase transition ending the supercooling. As expected, for all cases with non-negligible supercooling the peak height reaches a nearly constant value as in these cases the GW source at 
the PT is using already the entirety of the energy budget and eq.~\ref{eq:PTabundance} reaches its large $\alpha$ limit. 

The effect of supercooling on the string network is much more complex. This is because the inflationary period not only redshifts the previously generated spectrum according to the VOS description, but it also dilutes the network of strings. Depending on how much the network is diluted, it can take a significant amount of time for the density to grow back to the scaling solution~\cite{Cui:2019kkd}. It is only after that time that loops will be produced at the same rate as before the inflationary period and their decay into GWs, which is the main source for the spectrum, will also go back to the scaling result. Due to this fact, we see that longer periods of inflation shift to lower frequencies 
the return of the spectrum to its standard cosmology form (black solid lines) while the reheating temperature remains fixed. 
In fact since the return to scaling defines when the string spectrum goes back to its standard form there is a degeneracy between reheating and number of efolds. If we increase the number of efolds of thermal inflation increasing the dilution but also increase the reheating temperature to give strings more time to grow back into scaling at the same time their spectrum will remain unchanged.
In order to quantify this dependence we recall that the scaling of string energy when chopping becomes unimportant during inflation $\rho_\infty\propto a^{-2}$ while after inflation $\rho_\infty/ \rho_R \propto a^{2}$. This means the number of efolds spent by the network in inflation is the same as the number of efolds needed after it for the strings to grow back.
Using this fact and $a\propto T^{-1}$ we see that $N_e\propto \log T_{\rm reh}^{2}$ and changing the number of efolds according to this formula as we vary the reheating temperature we would always find the same GW spectrum.
Luckily this degeneracy in our scenario is broken by the phase transition spectra which are very insensitive to the exact length of the inflationary period while their frequency shifts with reheating temperature.

\begin{figure}[t]
    \centering
  
\includegraphics[width=0.72\linewidth]{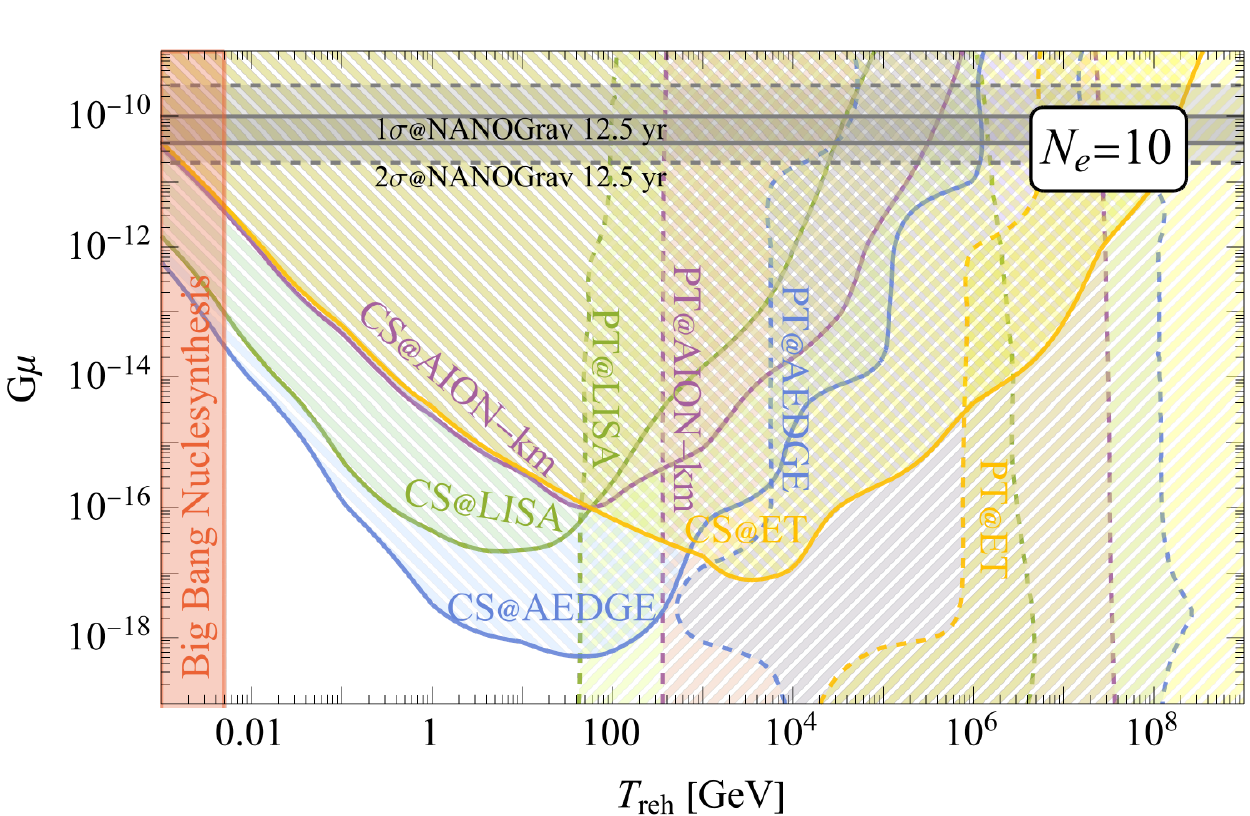}
\includegraphics[width=0.72\linewidth]{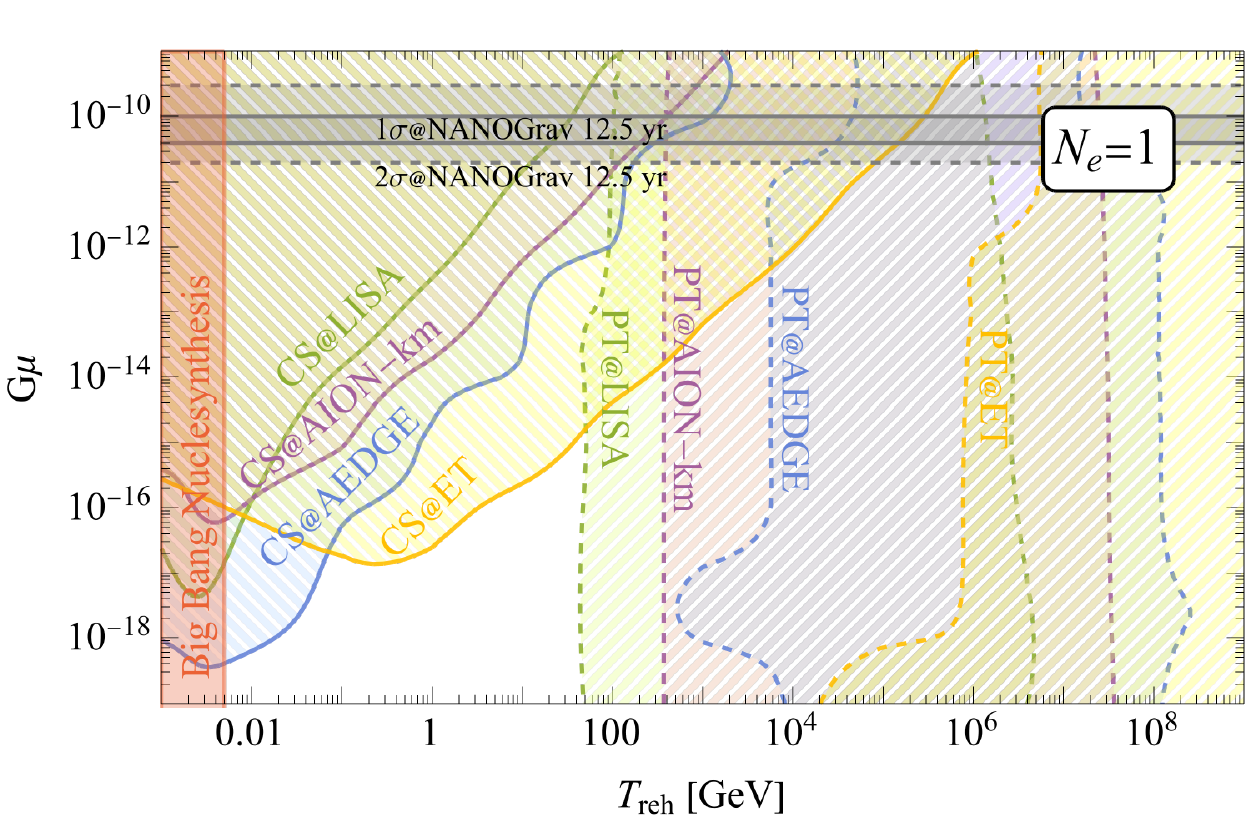}
\caption{\it \label{fig:SNRPlot}
\footnotesize Part of the parameter space where our modified signals will be observed with $SNR>10$. We separate the calculation into the modified cosmic string tail (solid contours) and the phase transition contributions (dashed lines). We take into account only modified parts of the spectrum where the amplitude is at least $10\%$ away from the standard string spectrum with a given $G\mu$.
For this example, we used $\beta/H=10$ for the PT, while the length of supercooling is taken to be $N_e=10$ (top panel) and $N_e=10$ (bottom panel).}
    
\end{figure}

Figure~\ref{fig:SNRPlot} shows parts of the parameter space in which our signals will be visible in upcoming experiments with $SNR>10$ (see eq.\eqref{eq:SNR}). Here we fix $\beta/H=10$  and vary the reheating temperature 
$T_{\text{reh}}$ as well as the tension of the string network $G\mu$.
The grey bands show the values giving a one and two sigma fit to the NANOGRav data~\cite{Ellis:2020ena}. The two panels show the dependence of the results on the amount of supercooling; we fix $N_e=10$ in the upper panel and 
$N_e=1$ in the lower one. 
The solid and dashed contours indicate where a given experiment would observe the PT spectrum and the modified cosmic string spectrum respectively. 
By modified spectrum here we mean one that can be distinguished from the result obtained assuming standard cosmological expansion. 
In practice, when computing the integral in~\eqref{eq:SNR} we include only the parts of the spectrum where the amplitude deviates from the standard one (black lines in fig.\ref{fig:SpectraPlot}) by at least $10\%$ as an example of a conservative threshold.

Regions, where both solid and dashed contours overlap in Fig.~\ref{fig:SNRPlot}, indicate where we would see a smoking gun signal of supercooling by observing both the PT spectrum and the modification of the cosmic string spectrum. 
The prospects grow with the amount of supercooling, since the increasingly 
diluted string network retains the modifications at lower frequencies 
that are radiated later as the network slowly comes back to scaling after 
dilution.

\label{sec:results}

\section{Conclusion}
\label{sec:conclusion}

Gravitational waves provide a pathway to test new high energy physics
phenomena that took place in the early pre-BBN stages of the evolution of the
universe. 
We considered a particularly well-motivated non-standard cosmological 
era known as supercooling or secondary inflation, and 
we studied its effects on the SGWB generated by a network of cosmic strings. 
Cosmic strings have a reasonably well understood GW emission spectrum. 
In particular, the emission during the standard radiation-dominated era results in a GW spectrum with a long flat plateau at high frequency. We can use this featureless spectrum to propagate through it various assumed cosmological histories and thus probe them with observational data.

In this paper, we have worked out in detail the structure imprinted 
on this SGWB by an epoch of supercooling. 
We find that such an early period 
of thermal inflation leaves a characteristic descending cosmic string plateau plus the peak produced at the end of supercooling.
This can be used to distinguish thermal inflation from any other mechanism that can suppress the SGWB from strings at high frequencies, such as an early matter domination period (see Figure~\ref{fig:SpectraPlot}). The reason is that supercooling will forcibly end in a first order phase transition, whereas the other contributions might not.

For typical phase transition parameters with $\beta/H=10$ involving large supercooling, we found that detection prospects depend crucially on the length of the supercooling period if we aim to find spectra from both sources. For a very short thermal inflation period with $N_e=1$ the combined data from future experiments will probe the features of the spectrum with string tension above $G\mu=10^{-15}$ and only if the reheating temperature is around $T_{reh}=100$ GeV. For larger string tensions close to the ones fitting the NANOGrav signal the detection range extends to $T_{reh}\in [10^2,10^5]$ GeV. For a long period of supercooling $N_e=10$ the detection prospects reach down to  $G\mu=10^{-18}$ and extend to $T_{reh}\in [10^2,10^8]$ GeV
for strong string spectra close to PTA limits (see Figure~\ref{fig:SNRPlot}).
Crucially, these include both the emission from the strings and 
phase transition itself which can be used as a smoking gun to distinguish an epoch of supercooling from other modifications of the early universe evolution such as an era of early matter domination.

\section*{Acknowledgements}
FF would like to thank the University of Warsaw for its hospitality. The work of FF was supported in 
part by the U.S. Department of Energy under Grant No. DE-SC0017987. 
This work was supported by the Polish National Science Center grants 2018/31/D/ST2/02048 and 2018/30/Q/ST9/00795, and the Polish National Agency for Academic Exchange within Polish Returns Programme under agreement PPN/PPO/2020/1/00013/U/00001.

\medskip



\bibliographystyle{JHEP}
\bibliography{main.bib}

\end{document}